\documentclass[iop]{emulateapj}
\pdfoutput=1

\usepackage{amsmath}
\usepackage{amssymb}
\usepackage{graphicx}
\usepackage{natbib}
\usepackage[breaklinks,colorlinks,citecolor=blue]{hyperref}
\usepackage{sidecap}
\usepackage{comment}
\usepackage[normalem]{ulem}

\newcommand{\WMAP}{\textsl{WMAP}}

\newcommand{\wmap}{{\WMAP}}
\newcommand{\Planck}{{\textsl{Planck}}}
\newcommand{\planck}{{\textsl{Planck}}}
\newcommand{\lcdm}{\ensuremath{\Lambda}CDM}

\renewcommand{\ell}{\ensuremath{l}}
\newcommand{\be}{\begin{equation}}
\newcommand{\ee}{\end{equation}}

\newcommand{\beq}{\begin{equation}}
\newcommand{\eeq}{\end{equation}}
\newcommand{\beqa}{\begin{eqnarray}}
\newcommand{\eeqa}{\end{eqnarray}}

\def\ba{\begin{eqnarray}}
\def\ea{\end{eqnarray}}

\newcommand{\barr}{\begin{array}}
\newcommand{\earr}{\end{array}}

\usepackage[flushleft]{threeparttable}
\usepackage{array}
\newcolumntype{C}[1]{>{\centering\let\newline\\\arraybackslash\hspace{0pt}}m{#1}}

\begin{document}

\title{An Examination of Galactic Polarization with Application to the Planck TB Correlation}
\author{J. L. Weiland, G. E. Addison, C. L. Bennett}
\affil{Department of Physics and Astronomy, Johns Hopkins University, 3400 N. Charles St, Baltimore, MD 21218, USA}
\email{jweilan2@jhu.edu}
\and
\author{M. Halpern, G. Hinshaw}
\affil{Department of Physics and Astronomy, University of British Columbia, Vancouver, BC, V6T 1Z1, Canada }

\slugcomment{accepted to ApJ, March 2020}

\begin{abstract}
Angular power spectra computed from \planck\ HFI 353~GHz intensity and polarization maps 
produce a \textsl{TB} correlation that can be approximated by a power law.
Whether the observed \textsl{TB} correlation is an induced systematic feature or a physical property
of Galactic dust emission is of interest both for cosmological and Galactic studies.
We investigate the large angular scale $E$- and $B$-mode morphology of microwave polarized thermal 
dust emission, and relate it to physical quantities of polarization angle and polarization fraction.  
We use empirical models of polarized dust to show that dust polarization angle is a key
factor in producing the \textsl{TB} correlation.  A small sample of both simulated and observed polarization
angle maps are combined with 353~GHz intensity and dust polarization fraction to produce a
suite of maps from which we compute \textsl{TB} and \textsl{EB}. Model realizations that produce a positive \textsl{TB} 
correlation are common 
and can result from large-scale ($>5^{\circ}$) structure in the polarization angle.
The \textsl{TB} correlation appears robust to introduction of individual intensity, polarization angle and polarization
fraction model components that are independent of the 353~GHz observations.
We conclude that the observed \textsl{TB} correlation is likely the result of large-scale Galactic dust polarization
properties.

\end{abstract}

\section{Introduction}

Cosmic microwave background (CMB) research is currently focused on obtaining high sensitivity
observations of the polarized CMB, as characterized by divergence free $B$-modes and curl-free
$E$-modes \citep{zaldarriaga/seljak:1997, kamionkowski/etal:1997}. $B$-modes are sensitive to the imprint of primordial gravitational waves sourced by inflation with an amplitude denoted by the tensor-to-scalar ratio parameter, r. Detection of these inflationary $B$-modes would constrain the energy scales of inflation and provide unique insight into the early universe.  $E$-modes observed on large spatial scales provide a direct constraint on the reionization
optical depth parameter $\tau$, and are important for cosmological neutrino mass constraints 
(e.g., \citealt{allison/etal:2015}). Current and upcoming experiments are targeting high multipole 
(up to $\ell\simeq3000-5000$) polarization measurements to more stringently test the \lcdm\ model and provide tighter constraints on the gravitational lensing of the CMB by large-scale structure \citep[e.g.,][]{benson/etal:2014,henderson/etal:2016,simonsobservatory:2019}.

Because the signal from the polarized CMB is subdominant to the Milky Way's polarized foreground emission,
it has become increasingly
important to both quantify and understand these foregrounds.
At microwave frequencies, the two significant emission mechanisms are synchrotron and thermal dust emission \citep{page/etal:2007,bennett/etal:2013,planck/10:2015,planck/04:2018}.
The polarized foregrounds are usually observed and characterized in terms of the Stokes parameters: $I$, $Q$, $U$, and $V$, but of primary interest are the linear polarization parameters $Q$ and $U$. While Stokes $Q$ and $U$ are observationally convenient, they are coordinate system dependent so they do not couple naturally to the underlying physics. For this, the $E$-mode and $B$-mode representation of orthogonal polarization components is ideal. Despite this, sky maps are usually shown in $Q$ and $U$ and foreground masking is generally implemented for $Q$ and $U$ maps before transforming and computing $E$ and $B$ power spectra.

\begin{figure*}[ht]
    \centering
    \includegraphics[width=6.5in]{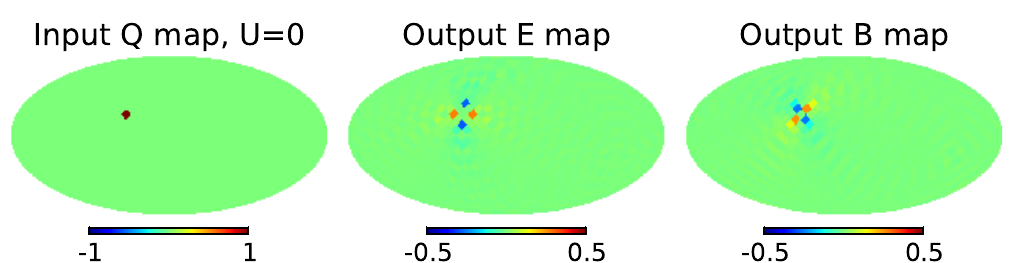}
    \caption{Illustration of how a single pixel with Stokes parameters $Q=1$, $U=0$ (left) transforms
    into $E$ (middle) and $B$ (right).  $E$ and $B$  may be expressed as second derivatives of $Q$ and $U$, hence a localized structure in
    $Q$ and $U$ is spread out in $E$ and $B$.   
    Note the change in color scale between $Q$ and the $E$ and $B$ maps.  
   As a consequence of non-locality, masks appropriate for [$Q, U$] data are not necessarily suitable for direct application to the corresponding  
   [$E,B$] data. 
   }
    \label{fig:q2eb}
\end{figure*}

A relatively recent foreground-related puzzle originates with analysis of \planck\ HFI data,
where the dominant polarized foreground is that from thermal dust emission.
Angular power spectra computed from HFI 353~GHz intensity{\footnote{When expressed in temperature units, the intensity map $I$ is often referred to as ``the temperature map'', or $T$.}} and polarization maps show a significant positive signal in the 
Temperature $B$-mode (\textsl{TB}) cross spectrum, but an \textsl{EB} cross spectrum consistent with zero signal \citep{planck/11:2018,planck/intermediate/54:2018, planck/intermediate/30:2016}. 
The \textsl{TB} spectrum is roughly consistent with a power law as a function of multipole moment $\ell$ for $\ell \lesssim 600$, although there is significant scatter in the data points about the fit.  We refer to this
characteristic as ``non-zero \textsl{TB}'' throughout this paper.
Physical arguments predict the \textsl{TB} and \textsl{EB} signals from the CMB to be statistically consistent with zero
(e.g., \citealt{zaldarriaga/seljak:1997}). Thus the two most likely sources for the non-zero \textsl{TB} signal 
are residual systematics in the data, and/or a non-zero \textsl{TB} contribution from Galactic thermal dust emission.

The reported detection of a non-zero \textsl{TB} signal in the HFI 353~GHz data is a finding of interest regardless of its origin, but carries different
ramifications depending on the source of that signal.  If the \textsl{TB} signal arises from
Galactic emission processes, it has implications for magnetic field structure studies and for CMB experiments.  Calibration of CMB observations may rely on hardware, astrophysical sources, and/or use
``self-calibration'', in which 
assumptions of null \textsl{EB} and sometimes \textsl{TB} spectra define the detector polarization angles and indicate the
presence of potential measurement systematics (e.g., \citealt{kaufman/etal:2014,koopman/etal:2016}).
Characterization of a non-zero \textsl{TB} sky signal would be essential knowledge for self-calibration \citep{abitbol/etal:2016}.
Efforts to find physical mechanisms to reproduce the \planck\ \textsl{TB} correlation are ongoing;
the model of \citet{filaments:2019} predicts both positive \textsl{TB} and \textsl{EB} (at a lower level)
based on alignment of diffuse ISM filaments with that of the Galactic magnetic field orientation.
Conversely, if the \planck\ 353~GHz \textsl{TB} signal 
arises from residual systematics in the data,  possible analysis bias in CMB as well as Galactic emission component studies may be 
introduced through the assumption that these maps are accurate templates of polarized Galactic dust emission.

This paper has two motivations: (1) obtaining some basic insights into the $E$ and $B$ morphology of polarized emission from Galactic foregrounds, and (2) attempting to understand the origin of the
non-zero \textsl{TB} signal in \planck\ data.  In Section~2, we discuss the [$Q,U$] to [$E,B$] transformation, and
show observations of polarized synchrotron and dust foregrounds in both representations in order to illustrate
relevant spatial features in the $E$ and $B$ maps.
The remainder of the analysis discussion (Sections 3, 4 and 5) is devoted to constructing models of $Q$
and $U$ Galactic dust maps assembled from a variety of intensity, polarization fraction and polarization angle
map components, and examining how the nature of these components affects both [$E,B$] and the resultant
\textsl{TB} power spectra.  Varying individual components within empirical models allows us to test the stability of the
observed non-zero \textsl{TB} spectrum.
Section~3 presents empirical models in which we examine the role of dust polarization angle and polarization 
fraction in the spatial structure of the [$E,B$] maps.  In Section~4, we compute \textsl{TB} and \textsl{EB} power
spectra of the models from Section~3, relate the origin of the non-zero \textsl{TB} spectrum to the
polarization angle, and compute \textsl{TB} spectra for additional models incorporating  
synthetic polarization angles.  
We discuss further tests of non-zero \textsl{TB} in Section~5, including intensity map variants and the addition of
\planck\ 217~GHz polarization data. 
Section~6 summarizes our conclusions.

\section{ Stokes [Q,U] and transformation to [E,B]}

$E$ and $B$ maps may be expressed as second derivatives of the Stokes $Q$ and $U$ maps (e.g., equation 9 of 
\citealt{zaldarriaga/seljak:1997}, see also \citealt{kamionkowski/etal:1997}).  As a result, the relation between the two systems in pixel space is non-local.  In Figure~\ref{fig:q2eb} we show the transformation of a single pixel with $Q=1$, $U=0$ into $E$ and $B$ maps.  The local $Q$ signal becomes a quadrupolar pattern in $E$ and $B$, where $B$ is rotated 45 degrees relative to $E$ \citep{hu/white:1997}.  In this example, the pixel scale sets the scale of the [$E,B$] structure, but more generally the Hessian of the [$Q,U$] maps determines the scale of structure in the [$E, B$] maps.  
As a consequence of non-locality,  masks appropriate for [$Q,U$] data are not necessarily suitable for the corresponding [$E,B$] data.
Similarly, note that simple characterizations of foreground structure in [$Q,U$], such as spectral indices, will not necessarily carry over to the corresponding [$E,B$] maps.

Throughout this paper, we utilize the {\textsl{healpy}}{\footnote{\url{https://github.com/healpy/healpy/}}} 
\citep{zonca/etal:2019} and {\textsl{HEALPix}}{\footnote{\url{http://healpix.sourceforge.net}}}
\citep{gorski/etal:2005}
analysis package.
The package function {\textsl{map2alm}} is used to transform $Q$ and $U$ using the 
spin-2 spherical harmonic basis functions with coefficients $a_{lm}^E$ and $a_{lm}^B$, and then  the function {\textsl{alm2map}} is used to evaluate $E$ and $B$ maps from the $a_{lm}$'s.  
The {\textsl{healpy}} transformation sets monopole and dipole ($\ell = 0,1$)
components to zero, so that all full-sky [$E,B$] maps that we show have zero mean and dipole.
Additionally, we retain the polarization convention adopted by \wmap\ and \planck\ data products, 
which differs from the IAU convention by the sign of Stokes $U${\footnote{\url{https://healpix.sourceforge.io/html/intro_HEALPix_conventions.htm}}}.
 
\begin{figure*} 
  \centering
  \begin{tabular}{cc}
    \includegraphics[width=3.5in]{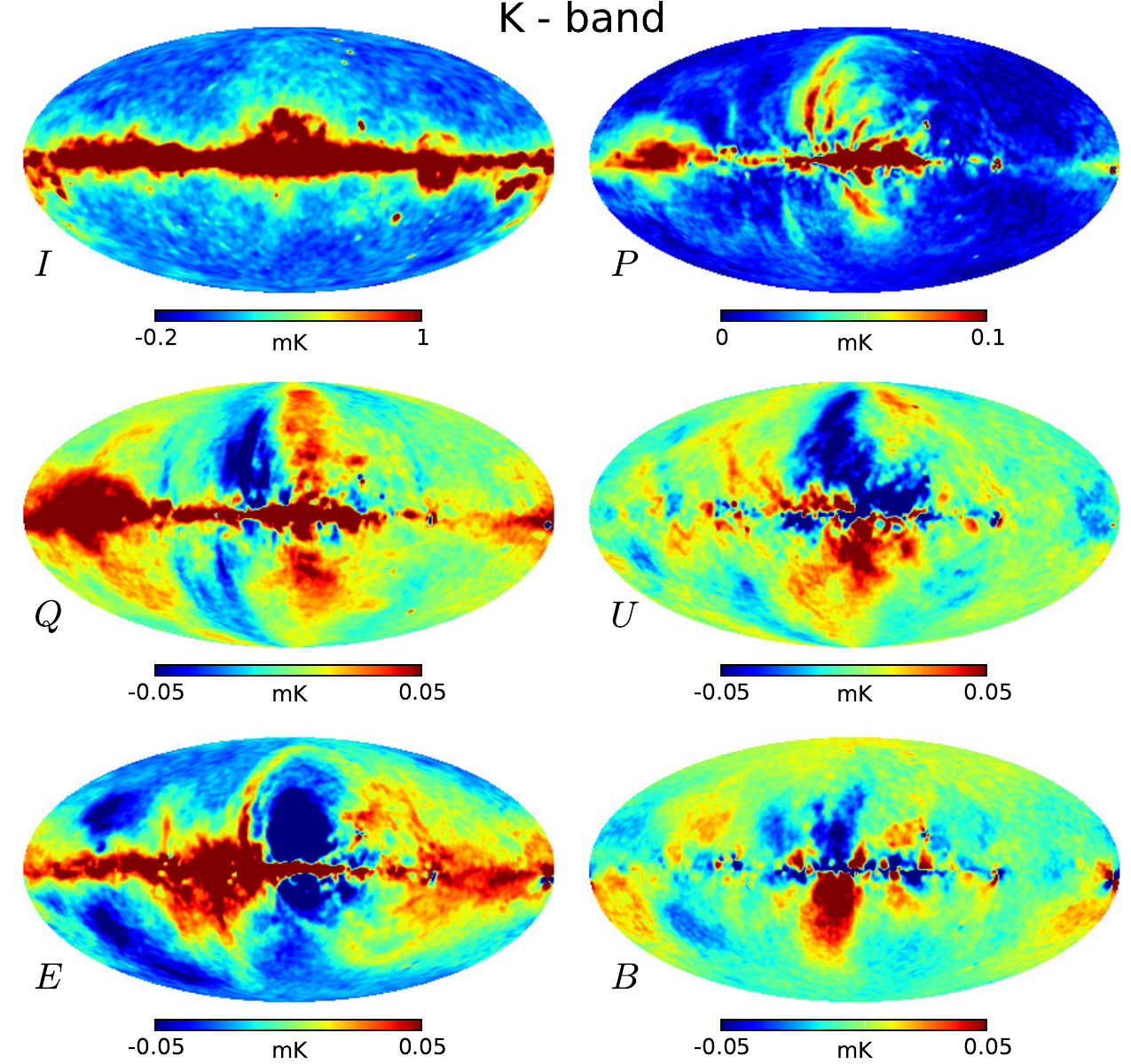}  
    \includegraphics[width=3.5in]{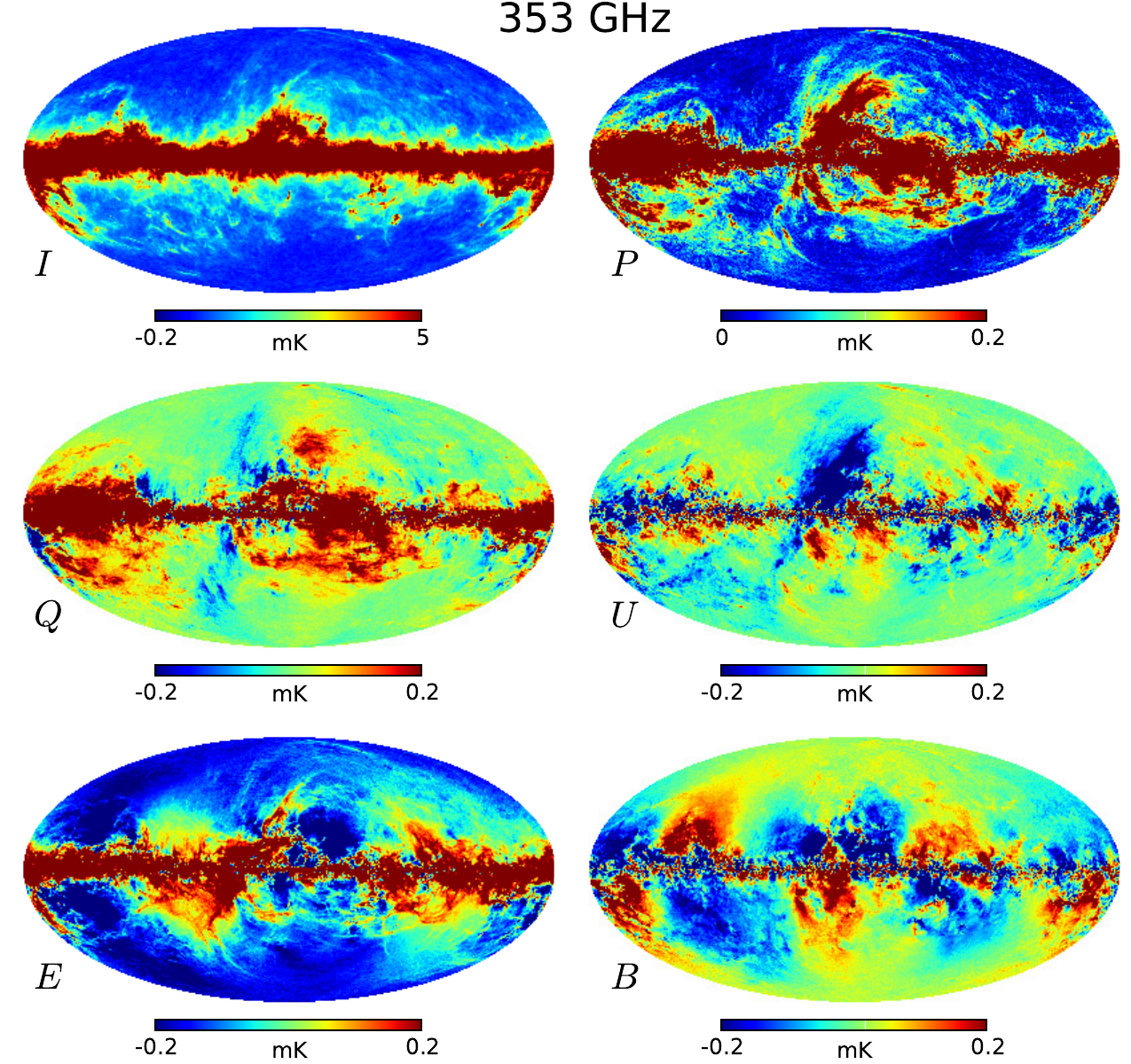}  
  \end{tabular}
  \caption{  Maps of unpolarized ($I$) and polarized ($P$) intensity at 23~GHz  (left half of plot, \WMAP\ K-band 9-year data) and 353~GHz 
  (right half, \planck/HFI 2018 data).  The bottom two rows show the polarization components in the form of $Q$ and $U$ maps (middle row)
  and $E$ and $B$ maps (bottom row).  The maps have been smoothed to reduce instrument noise.
  The CMB signal is negligible in the polarization maps, which are dominated by synchrotron emission at 23~GHz, and by thermal dust
  emission at 353~GHz.  Note that the $E$ maps of synchrotron and dust share many common features, whereas the $B$ maps appear to have less in 
  common.  In particular, the $B$-mode dust map exhibits a strong $\ell=3, |m|=3$ component.  We explore this further in Section 3.
 \label{fig:k353_eb}
 }
\end{figure*}

In Figure~\ref{fig:k353_eb}, we show the temperature (unpolarized intensity) map Stokes $I$, polarized intensity $P$ ($=[Q^2 + U^2]^{0.5}$, and the transformation from Stokes $Q$ and $U$ to $E$ and $B$ for \wmap\ K-band (23~GHz)
and \planck\ 353~GHz maps.  All maps in this paper are shown in Galactic coordinates, following astronomical
convention, with the Galactic center in the middle.
The K-band $Q$ and $U$ maps are dominated by Galactic synchrotron emission and
serve as a useful template for that foreground in polarization.  Similarly, the 353~GHz $Q$ and $U$ maps
may serve as templates describing polarized thermal dust emission.  The same unfortunately is not true
for the temperature maps, for while the 353~GHz $I$ map is dominated by thermal dust emission outside the
Galactic plane, the
K-band $I$ map represents a mix of CMB, synchrotron, free-free, and AME (anomalous microwave emission,
generally assumed to arise from spinning dust) components.  
Contributions also arise from point sources and cosmic infrared background asymmetries, but 
these are not critical to the larger angular scales discussed here.
Although we concentrate in this paper
on the more easily separable dust emission, we briefly discuss K-band in order to draw certain parallels, 
since both the synchrotron and dust emission are influenced by the same Galactic magnetic field.

The Galactic plane is clearly visible in all maps of Figure~\ref{fig:k353_eb}, but the details of the structures off the plane are not necessarily physically intuitive, particularly in polarization.  The signal
from polarized CMB is present in these maps, but is subdominant to Galactic components, such that high Galactic
latitude structure at these frequencies represents foreground behavior rather than that of the CMB. 
The $E$-mode maps for both K-band and 353~GHz show very similar large-scale structure.  The $B$-mode maps are
less alike, with the 353~GHz $B$-mode map showing a strong contribution from the 
$\ell=3, |m|=3$ spherical harmonic modes, evidenced as a regular +/- pattern alternating above and below the
Galactic plane. 

The above figures establish a general overview of the visual appearance of $E$ and $B$-mode
maps for polarized microwave foregrounds.
In the next section, we step through some simple models to gain insight into large angular scale $E$ and $B$ map structures.

\begin{figure*} 
\begin{center}
\includegraphics[trim={3cm 0 0 3cm},clip,width=6in]{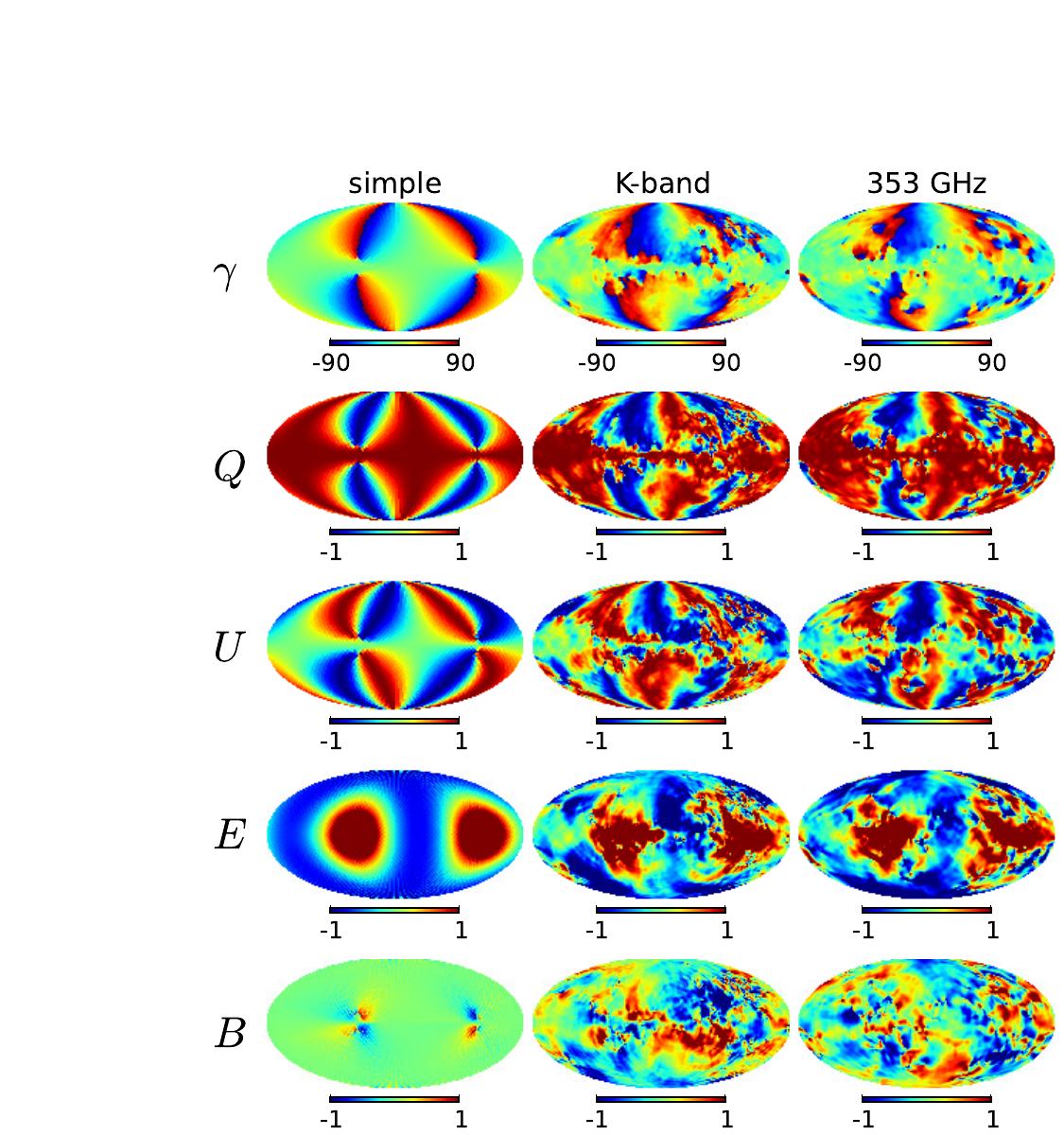}
\end{center} 
\vspace{-0.25cm}
\caption{Diagnostic polarization maps made to illustrate the role that polarization angle $\gamma$ plays in the morphology of [$Q,U$] and [$E,B$] maps.  In these maps we set the polarization intensity $P = If = 1$ and generate polarization maps from three different sources of polarization angle.  
Left column: $\gamma$ derived from a simple magnetic field model. Middle column: $\gamma$ derived from \WMAP\ K-band data, which mainly traces synchrotron emission. Right column: $\gamma$ derived from HFI 353~GHz data, which mainly traces dust emission.  The first row shows the source polarization angle, the next two show the $Q$ and $U$ maps induced by $\gamma$, and the last two show the induced $E$ and $B$ maps.  Comparison with the corresponding maps in Figure~2 shows that the two-lobed large-scale E-mode morphology can be attributed primarily to the polarization angle geometry, and in turn, the GMF. The B-mode map is not well predicted by $\gamma$ alone.
\label{fig:testfig4}
}
\end{figure*}

\section{Empirical Models}

Stokes $Q$ and $U$ maps may be constructed from an empirical model expressed in terms 
of a polarization angle $\gamma$ and polarization fraction $f$ as follows,

\begin{equation}
Q = I {f}  \cos (2\gamma)
\end{equation}
\begin{equation}
U = I {f}  \sin (2\gamma)
\end{equation}
where $I$ is the temperature map, such that
$f = P/I$ and $\gamma = 0.5~{\tan}^{-1}(U,Q)$.
We deal with models designed to represent only a single foreground emission component at a
given frequency.  

The polarization fraction is influenced by a number of factors. For example, a complex line-of-sight column containing multiple emission regions will result in a lower polarization fraction due to the decoherence of preferred directions (e.g., the magnetic field directions) and emission conditions along the column.  The polarization angle is influenced by the spiral structure imposed by
the global Galactic magnetic field (GMF) and more localized phenomena (e.g., MHD turbulence), and particle-field alignment properties. An observed temperature map $I$ is generally used to provide realistic small-scale information.

In this section, we construct models of [$Q,U$] in which we vary choices for $f$ and $\gamma$,
keeping $I$ constant, 
and note the effect of these individual parameters on the resultant [$E,B$] spatial structure.
We first examine the role that the polarization angle plays in the morphology of the [$Q,U$] and [$E,B$] maps, as illustrated in Figure~\ref{fig:testfig4}.
We start with three separate inputs of polarization angle, convert them via equations 1 and 2 to $Q$ and $U$ maps assuming $I$ and $f$
are everywhere unity over the sky, and then transform those [$Q,U$] maps to $E$ and $B$.  
Two of the three input polarization angle maps are constructed from observations discussed previously: \wmap\ K-band and \planck\ 353~GHz $Q$ and $U$ maps, smoothed to $4^\circ$ FWHM resolution.  The third polarization angle example is computed from a simple geometrical model of the GMF,
as parametrized by the logarithmic spiral arm model described in \cite{page/etal:2007}.  As discussed in Section 4 of that paper, we  adopt
a dust scale height of 100 pc, a spiral arm opening angle  $\phi_0 \simeq 25^\circ$ and tilt $\chi_0 \simeq 0^\circ$, with the polarization angle $\gamma$ 
as specified in their equation $11$.
 This representation, which we denote as ``simple $\gamma$'', provides a polarization angle map dominated by large angular scale morphology.  The polarization
angle as observed from the solar neighborhood does not represent the true magnetic field orientation as
seen by an outside observer, but is rather a projection.  The sign of $\gamma$ wraps at the directions
roughly corresponding the observer's view looking in both directions down the local Orion arm, roughly
at Galactic longitudes $65^{\circ}$ and $240^{\circ}$.

\begin{figure*} 
\begin{center}
\includegraphics[width=6.0in]{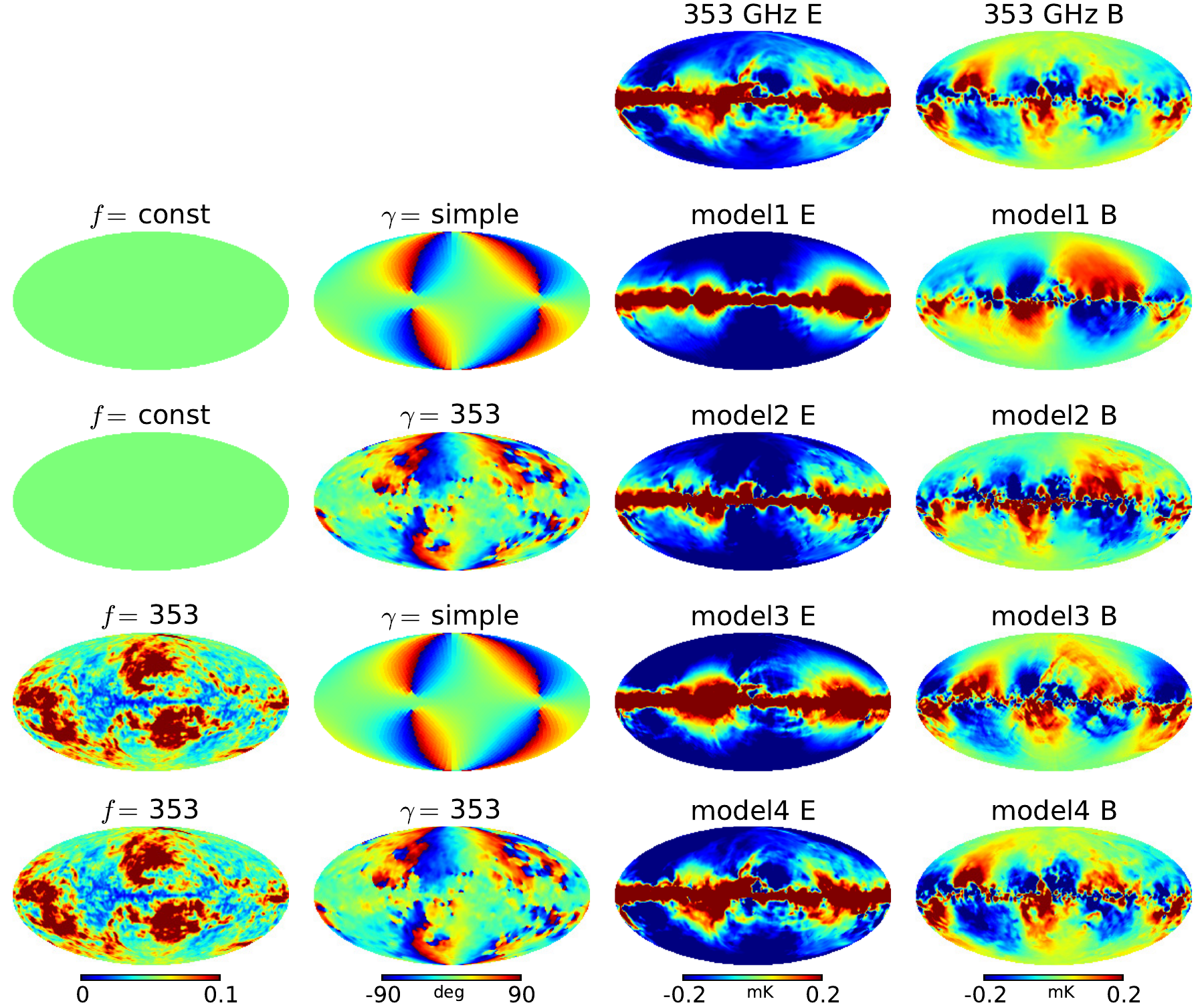}
\end{center} 
\vspace{-0.25cm}
\caption{Four empirical models of 353~GHz polarized dust emission are produced from the \planck\ 353~GHz temperature map 
and two separate variants each of polarization fraction $f$ and polarization angle $\gamma$ (see equations
1 and 2).  The variants of $f$ and $\gamma$ are chosen to span possible morphologies: $f$ as either 
spatially constant ($=0.05$) or a lightly smoothed version derived from \planck\ data, and $\gamma$ from either a simple GMF 
or computed from 353~GHz [$Q,U$]. The resultant [$E,B$] morphologies for each model are shown in 
rows 2 through 5, for
comparison to that of the data (first row). The first and second columns of each model row indicate
the choice of $f$ and $\gamma$.  The third and fourth columns present the [$E,B$] model maps derived
from those parameter choices.  As in Figure~\ref{fig:testfig4}, $\gamma$ is primarily responsible for the large-scale two-lobed $E$-mode morphology.
Obtaining the strong observed $\ell=3$ $B$-mode, however, requires a polarization fraction morphology 
similar to that of the 353~GHz data (rows 4 and 5).
\label{fig:testfig6}
}
\end{figure*}

We show each of these three polarization angle maps in the top row of Figure~\ref{fig:testfig4}. 
The next two rows illustrate the $\rm{cos}(2\gamma)$ and $\rm{sin}(2\gamma)$ geometrical patterns
that heavily influence the appearance of the $Q$ and $U$ maps shown in Figure~\ref{fig:k353_eb}.
The $E$ and $B$-mode maps resulting from these $Q$ and $U$ maps are shown in the fourth and fifth rows
respectively.  
All $E$-mode maps in this figure show two prominent large-scale emission lobes centered about the viewing directions of the Local Arm as defined by the GMF orientation. The lobe structure is
most obvious in the simple $\gamma$ $E$-mode map, but is strongly present both in the $\gamma =$
K-band and 353~GHz $E$ maps as well.
Comparing the $E$-mode
maps derived from the K-band and 353~GHz polarization angle to those of the data in Figure~\ref{fig:k353_eb}, it is clear that much of this
large-scale two-lobed structure is attributable to the polarization angle alone.

Comparison of the bottom row of Figure~\ref{fig:testfig4} with that of the data make it clear that
that the $B$-mode morphology is not sufficiently explained with polarization angle alone. Having examined the role of $\gamma$ in [$E,B$] structure, we now look at the role played by the polarization fraction $f$ via a set of four empirical models shown in Figure~\ref{fig:testfig6}.
The figure concentrates on empirical models of the dust emission at 353~GHz only,
because the maps of $I$ and $f$ for dust are more easily derived than those for K-band synchrotron, as
explained earlier.  We choose to vary the polarization fraction $f$ between two extremes: a constant
$f=0.05$ over the entire sky, and that produced from a lightly smoothed ($0.2^{\circ}$) map of $P$ and $I$
as observed by \planck. The $P$ map is produced from \planck\ half-mission maps in order to avoid introducing
noise bias.  In addition, we choose two bracketing representations of $\gamma$:
that from the simple geometric GMF model and 353~GHz data shown previously in Figure~\ref{fig:testfig4}.  We permute these $f$ and $\gamma$ choices
with a fixed thermal dust emission temperature map from \planck\ to form the four model combinations.
Our versions of $\gamma$ and $f$ independently derived from 353~GHz data are in good agreement with those shown and discussed in \citet{planck/12:2018}.
We also note that the exact choice of
353~GHz intensity map does not significantly
alter our results: for example, use of the
\planck\ Commander 353 GHz dust map 
\citep{planck/04:2018} rather than the
353 GHz $I$ map produces similar models.

In Figure~\ref{fig:testfig6} the $E$ and $B$ maps for
the 353~GHz data are shown in the top row and may be compared with the $E$ and $B$ model maps in the four rows
directly below.  The two leftmost columns specify the $f$ and $\gamma$ maps used to produce the $E$ and
$B$ models shown in each row.  
On very large scales, the dust $B$-mode structure takes the form of a quadrupole about the Local
Arm viewing directions near $l=65^\circ$ and $240^\circ$, and expresses itself as a strong
$\ell=3, |m|=3$ $B$-mode pattern about the Galactic plane.
The last two rows of the figure illustrate that the specific 353~GHz 
polarization fraction is necessary to most closely reproduce the observed $\ell=3, |m|=3$ $B$-mode pattern. There is strong geometrical suppression of the polarization fraction along the
spiral arm viewing directions, which is expected because of the complex astrophysics from
multiple emission regions along each line-of-sight.  

\begin{figure} 
  \centering
    \includegraphics[width=3.0in]{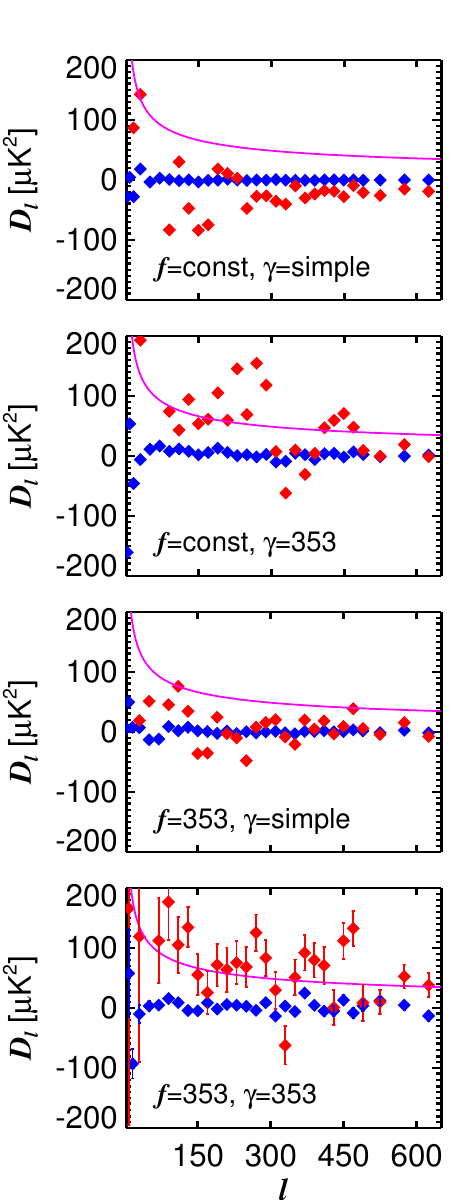}
  \caption{Binned \textsl{TB} (in red) and \textsl{EB} (in blue) power spectra computed for the four empirical models presented in Figure~\ref{fig:testfig6}, over $\approx 75$\% of the sky.  The magenta line shown in each panel represents the \textsl{TB} power-law fit found by \citet{planck/11:2018} for their largest sky fraction, and serves as a fiducial level for the observed positive \textsl{TB} signal.  Uncertainties estimated from the \planck\ FFP10 noise simulations are overplotted in the last panel; model uncertainties can be expected to be of this magnitude or less because of the reduced noise in some model components.
  A positive \textsl{TB} signal arises only for the two models formed using a polarization angle based on 
  353~GHz data.  We conclude that origin of the observed non-zero \textsl{TB} signal primarily lies with the
  $\gamma$ morphology.
  }\label{fig:polang_effect_on_tb}
\end{figure}

\section{\textsl{TB} and \textsl{EB} spectra of empirical models}

We use the four empirical models presented in the previous section to explore which geometrical
factors contribute most to producing the non-zero \textsl{TB} signal detected by \Planck.  
We construct half-mission $Q$ and $U$ maps using $I$, $f$ and $\gamma$ components at \textsl{HEALPix} $N_{side} =2048$ resolution
and evaluate the \textsl{TB} and \textsl{EB} power spectra using \textsl{PolSpice}\footnote{\url{http://www2.iap.fr/users/hivon/software/PolSpice/}}
\citep{szapudi/etal:2001, chon/etal:2004}.  
The resultant power spectra are shown in Figure~\ref{fig:polang_effect_on_tb}.
The use of half-mission splits is necessary
to avoid noise correlations in the power spectra when generating model components based on the 353~GHz data.
Because the $I$ component is always fixed to that of the 353~GHz data, a mask that excludes 
strong non-dust emission sources present in the data is used.  
In order to compare our results with that of \citet{planck/11:2018}, we adopt
the masking recipe described in \citet{planck/11:2018} for their analysis mask designated ``LR71''.
This mask maximizes analyzed sky area while excluding regions of strong 
CO line emission (primarily near the Galactic plane) and high-latitude polarized point sources above a certain threshold.   
Due to inexactitudes in the published masking specifications,
our mask is slightly different and permits roughly 75\% of the sky to be analyzed after apodization (rather than 71\%).
For visual clarity, power spectra are plotted using multipole bins; the binning ranges are as defined in Table~C.1
of \citet{planck/11:2018}.  Power spectra are expressed as $D_\ell = \ell(\ell+1) C_\ell / 2\pi$.

The bottom-most plot of Figure~\ref{fig:polang_effect_on_tb} shows binned \textsl{TB} (red) and \textsl{EB} (blue) 
produced from 353~GHz data half-mission cross-spectra, and represents the case where $f$ and $\gamma$ 
both arise from 353~GHz data.  This image illustrates independent confirmation of the non-zero \textsl{TB} 
signal and
\textsl{EB} spectrum roughly consistent with zero presented in Figure~6 of \citet{planck/11:2018}.  
\citet{planck/11:2018} estimate uncertainties for their \textsl{TB} and \textsl{EB} spectra based on analysis of
the full-focal plane simulations (FFP10) from the 2018 data release.  Similarly, the
uncertainties plotted in this panel are derived by applying the same analysis methods to the first
100 FFP10 simulations, which include noise, residual systematics, CMB and dust emission components.

\begin{figure*} 
\begin{center}
\includegraphics[width=6.0in]{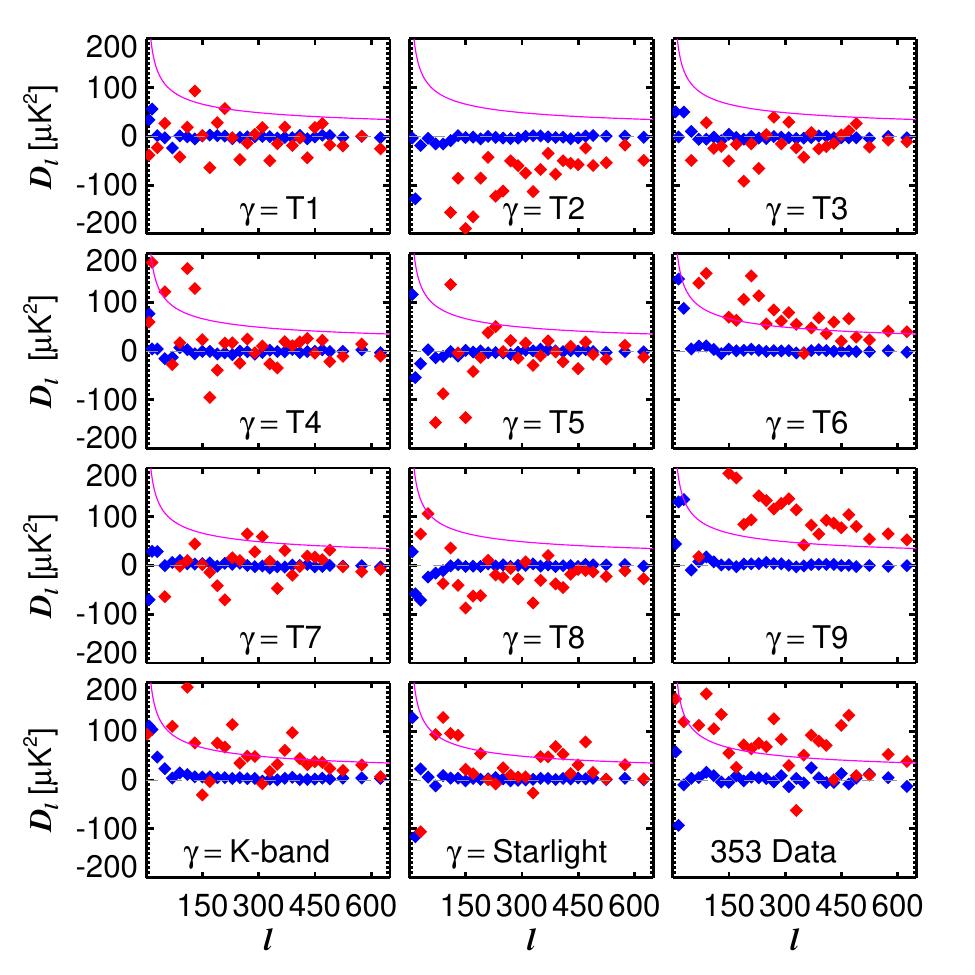}
\end{center} 
\vspace{-0.25cm}
\caption{ \textsl{TB} (red) and \textsl{EB} (blue) spectra of 353~GHz $I$, $Q$ and $U$ maps, computed for the 75\% mask. All panels, except the
lower right (``353 data"), show spectra derived from modeled $Q$ and $U$ realizations.   Each realization 
is constructed from the same set of 353~GHz intensity and polarization fraction maps based on \Planck\ data, but
each possesses a unique polarization angle morphology.  The nine panels labeled with the 'T1 - T9' prefix utilize 9 TIGRESS realizations of polarization angle corresponding to different observer locations within the solar neighborhood (see text). Also included are \textsl{TB} and \textsl{EB} spectra for two realizations built using data-based polarization angles which are independent of \planck\ observations. These are shown in the first two panels in the last row, and either use the \wmap\ K-band polarization angle (`K-band'), or $\gamma$ from the polarization of dust by starlight (`Starlight', \citealt{heiles:2000,page/etal:2007}).
The bottom right panel shows the \textsl{TB} and \textsl{EB} spectra for the 2018 \planck\  353~GHz data for comparison with the empirical models.  The magenta line shown in each panel represents the approximate \textsl{TB} power-law fit found by \citet{planck/11:2018} for their LR71 mask.
Although we show a small number of sample models, finding realizations which produce
a consistently positive \textsl{TB} power spectrum is not difficult.  The two models that used data-based
polarization angle maps consistently show a positive \textsl{TB} signal. Our results suggest that
the \textsl{TB} signal is likely a real feature of the Milky Way polarization structure. 
\label{fig:testfig10}
}
\end{figure*}

The approximate power-law fit derived by \citet{planck/11:2018}
for their LR71 mask is shown in magenta, which we evaluate using their equation 1,
$D_\ell^{\textsl{TB}} = A^{\textsl{TB}}($\ell$/80)^{(\alpha_{\textsl{TB}}+2)}$, with $\alpha_{\textsl{TB}} = -2.44$.
The amplitude $A^{\textsl{TB}}$ is not directly tabulated, but can be computed from the product of the \textsl{EE} amplitude and mean \textsl{TE/EE} amplitude ratios in
their Table~1, multiplied by the quoted \textsl{TB/TE} power ratio of $\sim0.1$.  
The magenta line serves as a fiducial when comparing with the other three model spectra.

The remaining panels in Figure~\ref{fig:polang_effect_on_tb} illustrate the effect on \textsl{TB} and \textsl{EB} of varying the polarization fraction and angle between extremes in complexity (simple vs. data).  \textsl{EB} is little affected except
for variations at low multipoles.  A positive \textsl{TB} signal approaching observed levels (represented by the same magenta 
line) is present only when a 
polarization angle $\gamma$ based on 353~GHz data is used.  Thus a $\gamma$ with more complex spatial 
structure than that of the simple GMF model is required to produce  the observed positive \textsl{TB}, 
although this does not indicate what particular property of $\gamma$ is the cause.

\begin{figure*}
\begin{center}
\includegraphics[height=5in]{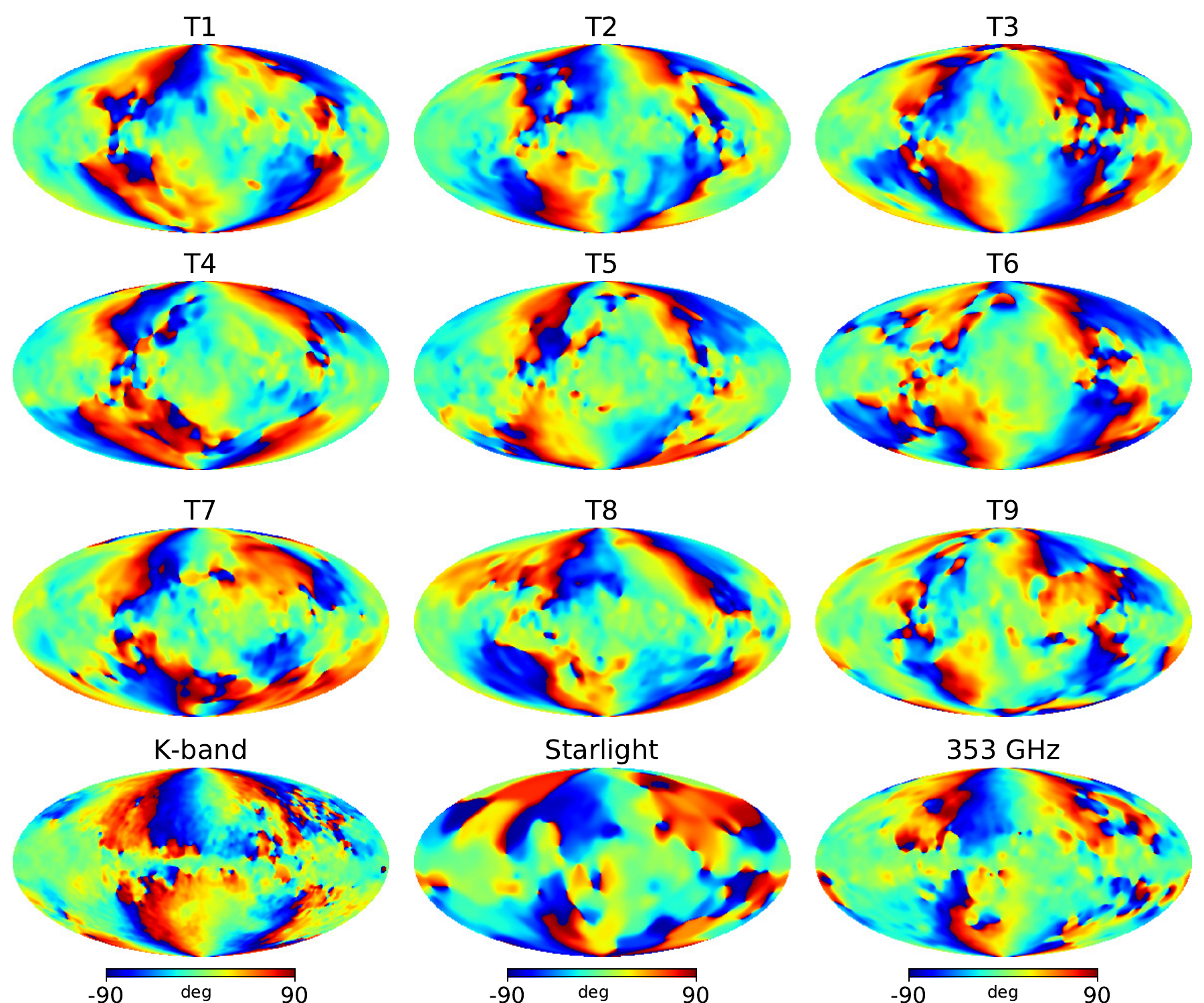}
\end{center} 
\vspace{-0.25cm}
\caption{Polarization angles used in forming the maps that produced the twelve \textsl{TB} and \textsl{EB} power
spectra shown in Figure~\ref{fig:testfig10}.
A non-zero \textsl{TB} spectrum can result from polarization angle morphologies which are
visually different from that of the 353~GHz data.  In Figure~\ref{fig:testfig10}, four of the 
eleven empirical models showed a positive \textsl{TB} signal.  Two of these are taken from data (`K-band' and
`Starlight') and strongly correlate with the 353~GHz $\gamma$ morphology.  However, realizations
T6 and T9 from the TIGRESS simulations bear little resemblance to the 353 $\gamma$ and yet produce
non-zero \textsl{TB}. 
\label{fig:testfig9}
}
\end{figure*}

\citet{planck/11:2018} note that current understanding of instrument polarization angle measurement errors preclude
a simple rotational angle error as the cause of the positive \textsl{TB} signal.  Additionally, a simple rotational
angle error should also leak into the \textsl{EB} spectrum, which is not observed.
Since it is apparent that $\gamma$ is the crucial factor in producing non-zero \textsl{TB}, we explore additional simple
models of $Q$ and $U$ in which we fix $I$ and $f$ components to match \Planck\ 353 GHz observations, 
but substitute in a variety
of simulated and data-based maps of $\gamma$ that are derived independently of the \planck\ observations.
This allows us to evaluate how easily one may obtain \textsl{TB} and \textsl{EB} dust spectra which are similar to those
observed by \planck\ at 353~GHz, but are not affected by possible residual systematics in the \Planck\ dust polarization 
angle. 

\begin{figure*}[ht]
    \centering
    \includegraphics[trim={1cm 0 0 0cm},clip,width=7in]{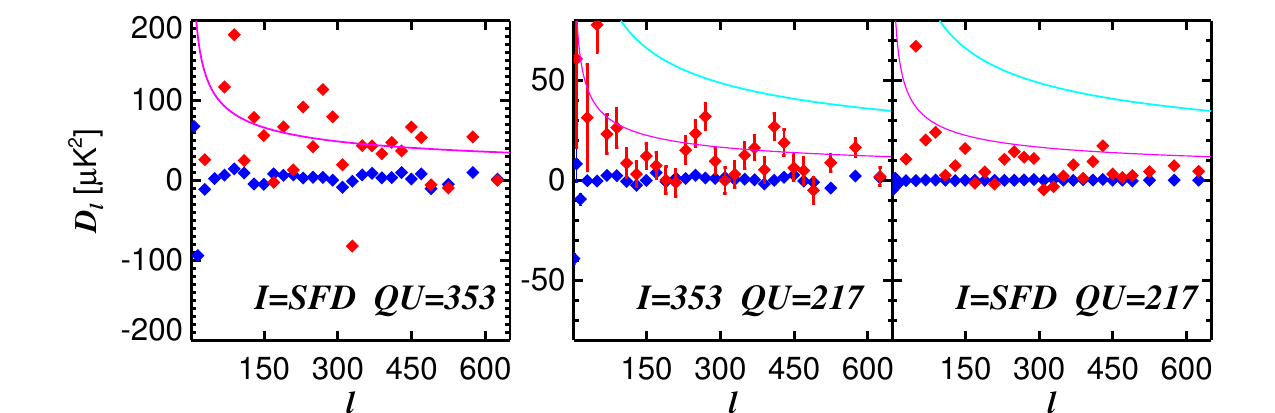}
    \caption{{\it{Left: }}Binned \textsl{TB} (red) and \textsl{EB} (blue) power spectra computed over the 
    75\% sky region for a simulation in which
    [$Q,U$] are the \planck\ 353~GHz maps, but $I$ is derived from the \citet{schlegel/finkbeiner/davis:1998} dust intensity model evaluated at the effective frequency of the 353~GHz band. There should be no common-mode systematics between the $I$ model and the 353~GHz
    data, yet a \textsl{TB} spectrum similar to that computed using only 353~GHz data (magenta line) results.
    {\it{Right, first panel: }} \textsl{TB} and \textsl{EB} cross power spectra computed from \planck\ 353~GHz half-mission 1 $I$ and 217~GHz
    half-mission 2 [$Q,U$] maps.  The cyan line represents the published 353x353 \textsl{TB} power spectrum; the magenta line is
    not a fit to the 217x353 spectrum but an extrapolation of the cyan line assuming a modified black body emission law. The 217x353 uncertainties are
    evaluated using the \planck\ FFP10 simulations.  {\it{Right, second panel: }}  Similar to preceding panel, but the 353~GHz intensity map now corresponds
    to that of the SFD 1998 model used in the leftmost panel.  This combination is independent of any \planck\ 353~GHz observations. 
    These tests support the idea that the 353~GHz positive \textsl{TB} spectrum arises from Galactic emission
    properties.}
    \label{fig:non353_croses}
\end{figure*}

Simulated $Q$ and $U$ maps of polarized dust emission have recently been made public\footnote{\url{https://lambda.gsfc.nasa.gov/simulation/tb_tigress_data.cfm}}
by \citet{kim/etal:2019}.
These ``TIGRESS sims'' provide a suite of polarized dust maps for nine observer positions within the solar neighborhood covering multiple evolutionary time-steps and are generated numerically using a multiphase, 
turbulent, magnetized interstellar medium (MHD code and TIGRESS ISM framework).  We create nine simulated
polarization angle maps from smoothed versions of the $Q$ and $U$ realizations for the nine observer positions 
and the $t= 360$ Myr time-step in their several-hundred Myr evolutionary interval.  In addition to these nine realizations, we
include two representations of polarization angle based on data.  The first of these is the \wmap\ K-band
polarization angle, which actually represents synchrotron emission, but has some similarities to the
dust polarization angle because both emission components are influenced by the same global magnetic field  
(synchrotron likely
follows gas with a larger scale height than the dust, and will trace a magnetic field that is
related but not identical to the field the dust experiences).
For our second data-oriented polarization angle map, we adopt the polarization angle constructed
from the \wmap\ dust polarization templates{\footnote{\url{https://lambda.gsfc.nasa.gov/product/map/dr5/templates_info.cfm}}}.  This angle is derived from observations of the optical polarization of starlight by dust grains \citep{page/etal:2007}, which may be expected to have some
similarities to the dust polarization angle at 353~GHz, although more sparsely sampled over the sky.  None of the eleven discussed options are available at high spatial resolution.  We replicate the pixels in $Q$ and $U$ maps to create maps pixelized at
HEALPix $N_{side}=2048$, but then smooth to $5^{\circ}$ FWHM resolution and compute the $\gamma$ map from the smoothed $Q$ and $U$ maps.
Thus all eleven of the $\gamma$ map realizations used in our models possess information only on $5^\circ$ scales or larger, but the $I$ and $f$ components of the models provide small-scale structure.

Figure~\ref{fig:testfig10} shows the \textsl{TB} and \textsl{EB} spectra computed from the 11 models of $Q$ and $U$ sky maps we constructed using the polarization angles described above.  We use the same 75\% sky mask described at the
beginning of this section.
Spectra are labeled according the the choice of
$\gamma$, since $I$ and $f$ are the same for all realizations, with ``T1-T9'' in the first three rows
referring to the nine TIGRESS simulations, and the last row showing the K-band and starlight polarization associated models, with the lower corner illustrating the 353 GHz data spectra (shown in Figure~\ref{fig:polang_effect_on_tb} with error bars) for comparison.
Among the 11 models, there is a wide variation in \textsl{TB} spectrum amplitude with the choice of 
polarization angle map (see the Appendix and Table 1 for estimated amplitudes).
Of the nine realizations that use the TIGRESS-derived polarization angle,
two (observers 6, 9) show \textsl{TB} spectra consistently above zero, and one (observer 2) is consistently below
zero. The two models that use
data-based $\gamma$ components both show a positive \textsl{TB} spectrum.  Although these latter two appear
on visual inspection to strongly correlate with the the 353~GHz polarization angle, the $\gamma$ maps 
corresponding to observers 6 and 9 (Figure~\ref{fig:testfig9}) do not.  
Although not shown, we also evaluated \textsl{TB} and \textsl{EB} for an additional nine TIGRESS-based models using the
maps from the last time-step in their evolutionary sequence.  In this case, three of the nine TB
spectra showed non-zero \textsl{TB} signals.

It therefore appears not to be difficult
to generate non-zero \textsl{TB} Galactic dust maps from simulations of the polarization angle constructed using current 
GMF models, and  non-zero \textsl{TB} seems robust to substitutions of $\gamma$ from other observational
data. We also note that spatial variations in polarization angle need not be small-scale to produce a positive \textsl{TB} signature, since the $\gamma$ maps here contain variations on $5^\circ$ scales or larger.

\section{Additional tests of \textsl{TB} signal persistence}

If non-zero \textsl{TB} is a property of Galactic emission, then it is reasonable to expect the effect not to
be confined to 353~GHz, nor to one specific mask.  
In this section, we briefly examine varying combinations of maps and masks that test for
persistence of the \textsl{TB} correlation.

\citet{planck/11:2018}  presented evidence for persistence of a positive 353~GHz \textsl{TB} correlation within the context of a
series of six nested masks  defined by dust emission intensity thresholds.
These masks admit between 24 - 71\% of the sky for analysis:  signal from higher Galactic latitude sky regions becomes more dominant 
as the analyzed sky area grows smaller, with a concomitant decrease in signal-to-noise. 
Their Figure  6 illustrates consistently positive \textsl{TB} power law amplitudes that progressively decrease in magnitude with sky fraction, 
indicating higher correlation closer to the Galactic plane.  As explained in Section 4, we have adopted a mask similar to their 71\% mask 
when computing the \textsl{TB} spectra presented in this paper.
However, we also explored an alternative masking strategy designed to test for the presence of non-zero \textsl{TB} in sky regions with the 
lowest expected contribution from \planck\ large-scale residual instrument systematics.  We used the FFP10 simulations 
to determine QQ and UU variances from noise and systematics in the polarization maps and adopted masking thresholds favoring regions
of low variance common to both Q and U.
Our most restrictive mask admitted $\sim$22\% of the sky for analysis, in two regions very roughly approximating truncated caps about the  ecliptic poles (the spatial truncation is due to the exclusion of CO emission near the Galactic plane described in Section 4).   
We found strong persistence of positive \textsl{TB} signal when using this mask, whereas a null detection would have argued for a possible non-sky origin. This is supportive, but not a definitive test, as it can only be performed for 353~GHz data because of signal-to-noise constraints, and 
the mask definition relies on some understanding of where the instrument systematics are strongest.

We next compute \textsl{TB} power spectra for cases in which the 353~GHz polarization maps are retained, but the intensity map is adopted from observations other than the \planck\ 353~GHz $I$ maps.
The use of an alternate $I$ map is designed to avoid possible leakage of 
instrument systematics between the intensity and polarization maps.
We tried combinations including \planck\ 545 and 857~GHz intensity maps scaled to 353~GHz emission
levels and the \planck\ dust intensity amplitude map derived from Commander multi-frequency analysis
\citet{planck/10:2015}, but still recovered the \textsl{TB} correlation at observed levels.
Because it is possible that there is some common-mode systematic between all \planck\ HFI frequencies,
we also evaluated a model for the 353~GHz dust intensity (\citealt{schlegel/finkbeiner/davis:1998}, SFD dust model 8). Although an older model, the SFD intensity map is based on IRAS and COBE/DIRBE data that
are completely independent of any \planck\ observations.  The \textsl{TB} spectrum obtained from the SFD 
intensity map and 353~GHz [$Q,U$] maps is shown in the left panel of Figure~\ref{fig:non353_croses}.
This recovered \textsl{TB} spectrum is similar to that for the 353~GHz data, as shown by the magenta line.

Finally, we attempted to examine dust polarization data from additional frequencies.  
This is a test of limited scope, because currently only \planck\ data provide the necessary sky coverage,
and dust signal-to-noise is decreasing at the next available frequency of 217~GHz.
Additionally, \textsl{TB} computation using only 217~GHz intensity and polarization data does not
produce a significant result in part because the 217~GHz $I$ dust signal is diluted by CMB at higher latitudes.  
We instead evaluate \textsl{TB} for the 75\% sky region
using the 353~GHz half-mission 1 $I$ map and 217~GHz half-mission 2 [$Q,U$] maps, shown in the
first of two plots in the double panel on the right of Figure~\ref{fig:non353_croses}.  In this panel, the cyan curve is the 353~GHz \textsl{TB}
power-law described previously.  The magenta line is the extrapolation of the observed 353~GHz
\textsl{TB} signal assuming the amplitude of the polarized dust emission follows a modified blackbody 
emission law described 
by a single physical dust temperature $T_d = 20 K$ and spectral index $\beta_d=1.6$.
There are more data points below the magenta line than above, so that the extrapolation is
not an exact predictor of the result.  However, data points for the 217x353 \textsl{TB} spectrum 
predominantly lie above zero, providing additional evidence for positive \textsl{TB} persistence beyond 353~GHz.
In the rightmost panel of Figure~\ref{fig:non353_croses}, we evaluate a 217x353 \textsl{TB} combination that does not include
any \planck\ 353~GHz data. In this combination, the intensity map is the same 353~GHz SFD dust model 8 mentioned previously,
and the polarization data are again the \planck\  217~GHz half-mission 2 [$Q,U$] maps.  Both right panels produce similar, and
positive, \textsl{TB} spectra (see Table~1 in the Appendix).

\section{Conclusions}

We have explored the basic morphology of $E$ and $B$ maps of Galactic dust emission and have used 
intensity, polarization fraction and polarization angle components to build
simple empirical models of the dust emission.  We computed \textsl{TB} and \textsl{EB} power spectra for these
models and investigated how $I$, $f$ and $\gamma$ contribute to \textsl{TB}.  We find that:

\begin{enumerate}
    \item Contributions to dust polarization map morphology from intensity ($I$), polarization angle 
    ($\gamma$) and polarization fraction ($f$) components are more easily distinguished in the $E$ and $B$ maps than $Q$ and $U$.  The $E$ component is strongly dependent on the large angular scale polarization angle structure, reflecting the GMF orientation.  
    Large-scale dust $B$-mode structure takes the form of a quadrupole about the Local
    Arm viewing directions near $l=65^\circ$ and $240^\circ$, and expresses itself as a strong
    $\ell=3, |m|=3$ $B$-mode pattern about the Galactic plane.  The exact morphology of
    the $\ell=3$ mode is constrained by the dust polarization fraction.
 
    \item Polarization angle is the key component in producing the observed \planck\ 353~GHz 
    non-zero \textsl{TB} spectrum.
   
    \item Intensity and polarization fraction maps with small scale structure  
    ($\ell > 600$), combined with large angular scale ($> 5^\circ $) polarization angle morphology, are capable of producing a non-zero \textsl{TB} spectrum similar to that of the \planck\ PR3 353~GHz data.  
    
    \item Within the context of our empirical models, we use a small sample of polarization angle realizations taken from the TIGRESS simulations to show that a non-zero \textsl{TB} spectrum is a property of individual realizations.  This implies a geometrical or spatial origin rather than a need for physics beyond that already incorporated in codes such as TIGRESS.
    Polarization angle maps derived from 
    observational data independent of \planck\ (\wmap\ K-band, dust polarization of starlight) also produce a non-zero \textsl{TB} spectrum.  
    
    \item The observed non-zero \textsl{TB} spectrum is persistent with variations of masking and alternate choices for the
    intensity map, including a dust intensity model completely independent of \planck\ observations.
    We find additional evidence for persistence of non-zero \textsl{TB} computed from 353~GHz $I$ and 217~GHz $[Q,U]$ maps.
\end{enumerate}

We have substituted estimates of $I$, $f$ and $\gamma$ components within our empirical models that
are independent of the \planck\ 353~GHz observations and found realizations for which 
non-zero \textsl{TB} persists.
Our findings support the conclusion that the \planck\ 353~GHz non-zero \textsl{TB} spectrum is a physical property
of Galactic dust polarization. 
In light of our studies in the paper, observing non-zero \textsl{TB} due to the Galactic dust foreground is not unusual or surprising.  It will be interesting to see if future experiments with higher
sensitivity detect non-zero \textsl{EB}.

\vspace*{0.15in}
This research was supported in part by NASA grants NNX16AF28G, NNX17AF34G, 80NSSC19K0526 and by the Canadian Institute for Advanced Research (CIFAR).  
This research has made use of NASA's Astrophysics Data System Bibliographic Services. 
Some of the results in this paper have been derived using the \textsl{healpy} and \textsl{HEALPix} package.
We acknowledge the use of the 
Legacy Archive for Microwave Background Data Analysis (LAMBDA), part of the High Energy Astrophysics Science Archive Center (HEASARC). 
HEASARC/LAMBDA is a service of the Astrophysics Science Division at the NASA Goddard Space Flight Center.  
We also acknowledge use of the \Planck\ Legacy Archive. \Planck\ is an ESA science mission with instruments and contributions 
directly funded by ESA Member States, NASA, and Canada.

\appendix
\section{Summary Table of \textsl{TB} Correlation Results}
\label{sec:summary_table}

In Table~1, we provide a summary of the components comprising the simple empirical models and data combinations 
that were used to create the \textsl{TB} spectra presented in Figures 4, 6 and 8.  In Section 6, we summarize how these tests
informed our conclusion that the observed \textsl{TB} correlation is likely the result of large-scale Galactic dust polarization properties. 
Also in Table~1, we provide an approximate
measure of the strength of the \textsl{TB} signal in each of the example spectra shown in these figures. 
We could not develop a
rigorous statistical representation of the binned \textsl{TB} power spectrum uncertainties for many of our empirical realizations, because each realization represents a combination of \planck\ data  with either noiseless models or components from other missions.   
As an approximation, we assume the statistical uncertainties in each bin  are roughly equivalent to those that are derived from the \Planck\ FFP10 simulations for the relevant \Planck\ frequencies (see Section 4).  We adopt these uncertainties and  perform a weighted least squares fit of a power law to the binned \textsl{TB} spectrum, where the power law takes the form
$D_\ell^{\textsl{TB}} = A^{\textsl{TB}}($\ell$/80)^{(\alpha_{\textsl{TB}}+2)}$, with fixed $\alpha_{\textsl{TB}} = -2.44$.
In this case, the only free parameter is the amplitude $ A^{\textsl{TB}}$, which may be taken as a measure of \textsl{TB} strength.
This is the same power law parametrization adopted by the \Planck\ Collaboration in presenting their original findings, and thus provides a common metric.
In the last column of Table~1, we tabulate $A^{\textsl{TB}}$ from the weighted fit to the binned spectrum.   The derived statistical
uncertainty for this parameter, assuming these approximations, is indicated in parentheses.

{\renewcommand{\arraystretch}{1.2} 

\begin{table}
\centering
\caption{Summary of \textsl{TB} Correlation Test Results}

\label{tab:summary_tab}
\begin{threeparttable}

\begin{tabular}{|l|c|c|c|r|}
\hline
\hline
\multicolumn{5}{c}{ Section 4: 353 GHz Empirical Models (varying components)}\\
\hline
\hline
Description & Model  & Polarization fraction, $f$ & Polarization angle, $\gamma$ & TB Amplitude{\tnote{a}} \\
\hline
Fig 4: I fixed, vary $f$, $\gamma$ components & $1$ &       constant  &  simple      &	   $-54\ (10)$ \\
                                                                           &  $2$ &      constant  & \Planck\ 353 &   $90\ (10)$ \\                       
                                                                           &  $3$ &      \Planck\ 353  & simple    &   $  2\  (10)$ \\                       
                                                                           &  $4$ &      \Planck\ 353  & \Planck\ 353    &   $  110 \ (10)$ \\                       
\hline
Fig 6: I and $f$   fixed, vary $\gamma$ only      & $1$  &      \Planck\ 353  &  TIGRESS 1     &	   $-12\ (10)$ \\
                                                                           &  $2$ &      \Planck\ 353  &  TIGRESS 2    &   $-156\ (10)$ \\                       
                                                                           &  $3$ &      \Planck\ 353  &  TIGRESS 3     &   $ -29\  (10)$ \\                       
                                                                           &  $4$ &      \Planck\ 353  &  TIGRESS 4    &   $  17 \ (10)$ \\                       
                                                                           &  $5$ &      \Planck\ 353  &  TIGRESS 5    &   $  -6\ (10)$ \\                       
                                                                           &  $6$ &      \Planck\ 353  &  TIGRESS 6    &   $  141\ (10)$ \\                       
                                                                           &  $7$ &      \Planck\ 353  &  TIGRESS 7    &   $     8 \ (10)$ \\                       
                                                                           &  $8$ &      \Planck\ 353  &  TIGRESS 8    &   $  -51\ (10)$ \\                       
                                                                           &  $9$ &      \Planck\ 353  &  TIGRESS 9    &   $  218 \ (10)$ \\                       
                                                                           &  $10$ &    \Planck\ 353  &  K-band          &   $  94 \ (10)$ \\                       
                                                                           &  $11$ &    \Planck\ 353  &  Starlight        &   $  59 \ (10)$ \\                       
\hline
\hline
\multicolumn{5}{c}{Section 5:  IQU Combinations that avoid use of only \planck\ 353~GHz data}\\
\hline
\hline
Description & Model  & Intensity Map & Q and U Maps & TB Amplitude{\tnote{a}} \\
\hline
Fig 8: Different choices for IQU Maps  & $1$ &       SFD 353  &  \Planck\ 353      &	   $ 73\ (10)$ \\
                                                             &  $2$ &      \Planck\ 353  &  \Planck\ 217   &   $ 19\ (2)$ \\   
                                                             &  $3$ &      SFD  353  &  \Planck\ 217   &   $ 13\ (2)$ \\   
                                                                                 
\hline
\end{tabular}

\begin{tablenotes}
\item[a]{ As described in the Appendix, this is $A^{\textsl{TB}}$ in units of $\mu$K$^2$ as derived from a power law fit to the binned \textsl{TB} spectrum
computed for each model.
 Uncertainties per bin are estimated from FFP10 simulations including noise, residual systematics, dust emission and CMB.  Approximate uncertainties for $A^{\textsl{TB}}$ are provided in parentheses.}
\end{tablenotes}
\end{threeparttable}
\vspace*{0.10in}
\end{table}

{\renewcommand{\arraystretch}{1.0}

\end{document}